\magnification=\magstep1
\advance\voffset by 3truemm   
\advance\hoffset by -1truemm   
\vsize=23truecm  \hsize=16.5truecm
\overfullrule=0pt
\hfuzz 15truept
\parskip=5pt
\baselineskip=12pt
\font\tit=cmbx10 scaled \magstep2
\font\subt=cmbx10 scaled \magstep1
\font\ssubt=cmbx10
\font\auth=cmr10 scaled \magstep1
\font\titabs=cmti9
\font\abs=cmr9
\def\title#1{\null\vskip 26truemm \noindent {\tit #1} \vskip 12truemm 
              \noindent {\auth By} } 
\def\authors#1{\noindent {\auth #1}   \vskip 5truemm}
\def\address#1{\noindent #1\vskip 7truemm}
\def\abstract#1{\vskip 19truemm\noindent 
   {\abs {\titabs Abstract} #1  }} 
\def\section#1{\vskip 12truemm \noindent{\tit #1}\vskip 5mm\noindent}
\def\subsection#1{\vskip 8truemm \noindent{\subt #1}\vskip 3truemm\noindent}
\def\subsubsection#1{\vskip 5truemm \noindent{\ssubt #1}\vskip 2truemm\noindent}


\def\lsim{\lower2pt\hbox{$\buildrel{<}\over{\sim}$}}
\def\gsim{\lower2pt\hbox{$\buildrel{>}\over{\sim}$}}
\def\rr{{\rm r}}\def\cc{{\rm c}}\def\dd{{\rm d}}
\title{Can Doppler Peaks discriminate among Inflationary models and Topological
Defect scenarios ?}
\authors{Mairi Sakellariadou}
\address{D\'epartement de Physique Th\'eorique, Universit\'e de Gen\`eve,
	 24 quai Ernest-Ansermet, CH-1211 Gen\`eve 4, Switzerland}
\abstract { Doppler peaks in the cosmic microwave background may allow us to 
distinguish among the two classes of theories---inflationary models and
topological defect scenarios---which attempt to explain the origin of 
structure formation in the universe.  We consider density perturbations
seeded by global textures in a universe dominated by cold dark matter.  We 
calculate the height and the position of the primary peak and conclude a 
different signature than the one obtained if the initial perturbations were 
due to the amplification of quantum fluctuations of a scalar field during a 
generic inflationary era. We believe that our analysis holds for all kinds of 
global defects and general global scalar fields. 
We then question the validity of the temporal coherence of the sources, 
assumed in the texture models. We finally discuss the temporal coherence of
cosmic string sources, through correlations of the energy and momentum in an
evolving cosmic string network in Minkowski space.}

\section
\noindent
One of the most important issues of modern cosmology is
the origin of the large-scale structure.  We believe 
that it was produced by gravitational instability
from small primordial fluctuations in the energy density, generated in the
early universe.  Within this framework there are two classes of 
theories to explain the origin of the primordial density perturbations. They 
can be due to quantum fluctuations of a scalar field during an 
inflationary era, or they may be seeded by topological defects 
produced during a symmetry breaking phase transition.  
Inflationary fluctuations  lead to an approximately scale-invariant 
(Harrison-Zel'dovich) spectrum of density perturbations, generated through a 
linear mechanism, with a Gaussian distribution of amplitudes on scales which 
are cosmological today.  These are {\sl passive coherent} 
fluctuations.  Topological defect perturbations lead also to an approximately 
scale-invariant spectrum of density perturbations, however
generated via a non-linear process, with constant amplitude on each scale at 
horizon crossing at all times.  These are {\sl active incoherent} 
fluctuations, for which causality requires the existence of a large-scale 
radiation white noise-spectrum.  Either of these two classes of theories 
predicts precise fingerprints in the cosmic microwave background (CMB) 
anisotropies, which can be used to differentiate among them using a 
purely linear analysis. 

The CMB fluctuation spectrum is usually parametrized 
in terms of multiple moments $C_\ell$, defined as the coefficients in the 
expansion of the temperature autocorrelation function
$$ \left\langle{\delta T\over T}({\bf n}){\delta T\over T}({\bf n}')
   \right\rangle\bigg|_{({\bf n\cdot n}'=\cos\vartheta)} =
   {1\over 4\pi}\sum_\ell(2\ell+1)\,C_\ell P_\ell(\cos\vartheta)\;, \eqno(1)
$$
which compares points in the sky separated by an angle $\vartheta$. 
The main physical mechanisms which contribute to the redshift of 
photons propagating in a perturbed Friedmann geometry are:
fluctuations in the gravitational potential on the last-scattering surface
({\sl Sachs-Wolfe} effect), acting on large angular scales ($\ell \lsim 50$)
\def\lsim{\lower2pt\hbox{$\buildrel{<}\over{\sim}$}}; acoustic waves in the 
baryon-radiation fluid prior to recombination ({\sl Doppler peaks}), acting on 
angular scales ($0.1^\circ\lsim \ \theta \lsim 2^\circ$); and 
suppression of CMB anisotropies due to the finite thickness of the
recombination shell as well as to photon diffusion during recombination
({\sl Silk damping}), acting on the smallest angular scales ($\ell\gsim  
1000$). Both, generic inflationary models and topological defect scenarios 
predict an approximately scale-invariant spectrum of density perturbations on 
large angular scales.  Thus CMB anisotropies on intermediate 
and small angular scales are very important.  If the two families of models
predict different characteristics for the Doppler peaks, one can discriminate 
among them. Inflationary perturbations predict coherent oscillations, with the 
primary Doppler peak at $\ell \sim 200$, having an 
amplitude 
$\sim$ 4--6 times the Sachs-Wolfe plateau, and the appearance of secondary 
oscillations [1].

\vskip 0.5truecm
To study the characteristics of the Doppler peaks in the CMB produced from 
textures, we will employ a gauge-invariant linear perturbation analysis.  
Neglecting the integrated Sachs-Wolfe 
(ISW) effect, the Silk damping and the contribution of neutrino 
fluctuations, the Doppler contribution to the CMB anisotropies is [2]
$$
\left[{\delta T \over T}({\bf x,n})\right]^{\rm Doppler} \approx 
{1\over 4}D_\rr({\bf x}_{\rm rec},\eta_{\rm rec}) + 
{\bf V}({\bf x}_{\rm rec},\eta_{\rm rec})\cdot {\bf n}\ , \eqno(2)
$$
where ${\bf V}$ is the peculiar velocity of the baryon fluid with 
respect to the overall Friedmann expansion, $D_r$ is a gauge-invariant 
variable describing the density fluctuation in the coupled baryon 
radiation fluid and  
${\bf x}_{\rm rec} = {\bf x} - {\bf n}\,\eta_0 $ (${\bf n}$ denotes
a direction in the sky, $\eta$ is conformal time, with $\eta_0$, 
$\eta_{\rm rec}$ the present time and the time of recombination respectively).
  
We study a two-component fluid system: baryons plus radiation, which prior
to recombination are tightly coupled, and CDM.  The evolution for the 
perturbation variables $D$ (density perturbation) and $V$ (velocity 
perturbation) in a flat background is given by
$${\cal D}\left(\matrix{D_\rr\cr
	                 D_\cc\cr}\right)
	=  S \ ,\eqno(3)$$
where subscripts ``r'' and ``c'' denote the baryon-radiation plasma and CDM, 
respectively.  In Eq.\ (3), ${\cal D}$ stands for a second order differential 
operator and $S$ denotes the source term, in general  
given by $S=4\pi G a^2(\rho +3p)^{\rm seed}$; in our case, where the seed is 
described by a global scalar field $\phi$, the source term is 
$S=8\pi G (\phi ')^2$.
Numerical simulations show that the average of $|\phi '|^2$
over a shell of radius $k$ can be modeled by [3]:
$\langle|\phi'|^2\rangle (k, \eta) =  
0.5 A{\tilde{\eta}}^2 \eta^{-1/2}[1+\alpha (k \eta)
+\beta (k \eta)^2]^{-1}, $
where ${\tilde{\eta}}$ is the symmetry breaking scale of the phase 
transition leading to texture formation; $A, \alpha, \beta$ are 
parameters of order 1.  For a given scale $k$, we chose the initial time 
such that the perturbation is super-horizon and the universe is radiation 
dominated.  With these initial conditions we solve the system of second order
equations for the perturbation variables, obtaining 
$D_\rr$ and $D_\rr'$.
The Doppler contribution to the CMB anisotropies is given by
$$
C_{\ell} = {2\over \pi} \int \dd k \left[{k^2\over 16} \big|D_\rr
(k,\eta_{\rm rec})\big|^2 j_{\ell}^2(k\eta_0) + {1\over(1+w)^2} \big|D_\rr'
(k,\eta_{\rm rec})\big|^2 (j_{\ell}'(k\eta_0))^2\right], \eqno(4)
$$
where $w=p_\rr/\rho_\rr$; $j_{\ell}$ is the spherical Bessel function of 
order $\ell$, and $j_{\ell}'$ its first derivative.  The angular power 
spectrum, shown in the figure, yields the Doppler peaks; we show 
separately the contribution of $D_\rr$ (upper dotted line), $D_\rr'$ (lower
dotted line), as well as their sum (solid line).  
\vskip 8.6 truecm
\midinsert\narrower
\midinsert\narrower 
\midinsert\narrower
\baselineskip=12pt
\noindent
Fig. The angular power spectrum for the Doppler contribution to the CMB 
anisotropies
is shown in units of $\epsilon$.  We choose the cosmological parameters
$h=1/2,\ \Omega_B=0.05$ and $z_{rec}=1100$.
\endinsert
\endinsert
\endinsert
The ISW effect will shift the
position of the first peak to somewhat larger scales, lowering
$\ell_{\rm peak}$ by (5--10)\% and possibly increasing slightly its amplitude
(by less than 30\%).  So the primary peak is displaced by 
$\Delta\ell\sim 150$  towards smaller angular scales than in standard 
inflationary models.  Silk damping will decrease the 
relative amplitude of the third peak with respect to the second one; however 
it will not affect substantially the height of the first peak, which is
$\ell(\ell+1)\,C_{\ell}|_{_{\rm Doppler}}=(2-3)~ 6C_2$ ~ [2]. 

\vskip 0.5truecm

Here, as well as in [4], we assumed maximum coherence for the texture models
and found that the peaks were preserved.  As emphasised in [5], the 
distinctive appearance  of Doppler peaks and troughs seen in inflationary 
calculations and texture models depend sensitively on the temporal coherence
of the sources.  Assuming little coherence the peaks are washed out, while 
an assumption of total coherence preserves them.  
An incoherent defect perturbation is effectively coherent and displays 
secondary oscillations, if the defect scaling coherence time is much bigger
than $2\pi \eta/ \xi_\cc \ $, where $\xi_\cc$ is the defect coherence length 
[6].  Assumning effective coherence for textures means that the coherence 
function 
$${\cal C}_{\Phi}(k\eta,k\eta ')\equiv {<\Phi(k,\eta)\ \Phi(k,\eta ')>
\over \sigma(\Phi(k,\eta))\ \sigma(\Phi(k,\eta '))},$$ 
is equal to 1, where 
$\sigma$ denotes the square root of the power spectrum.  Checking numerically
whether the unequal time corrrelator for $|\phi '|^2$ has an exponential decay
on a timescale which will define the coherence time, we conclude [7] that even 
though effective coherence is not fully justified for textures, 
the characteristic features of the first Doppler peak found
here, do indeed hold, while secondary oscillations should exist but be 
softened than the ones predicted according the coherent approximation.

\vskip 0.5truecm
We now question the validity of the coherence assumption for local gauge 
strings, since understanding the temporal coherence of string sources is 
very important when calculating their microwave background signals. 
The authors in [5] assumed that strings were effectively incoherent and 
obtained a rather featureless CMB power spectrum at large multipole $\ell$.
In [6] this assumption was justified by a numerical study of the two-time
energy density correlator.  The authors concluded the absence of secondary
oscillations and the validity of the totally incoherent approximation. 
Performing numerical experiments, we investigate scaling properties of the
power spectra and correlations of the energy and momentum in an evolving 
string network in Minkowski space [8] and measure the coherence time in the 
network.  We expect a network of cosmic strings evolving in Minkowski space
to have all the essential features of one in a Friedmann background, while 
the big advantage of Minkowski space is that the network evolution is very 
easy to simulate numerically.  To a good approximation, the cosmic string 
network can be thought of as consisting of randomly 
placed segments of string, of length $\xi/\sqrt{1-\bar v^2}$ and number
density $\xi^{-3}$, with random velocities;
$\xi$ denotes the energy density scale defined by $\xi^2=\mu/\rho_{inf}$ 
($\mu$ is the linear mass density and $\rho_{inf}$ is the density of 
string with energy greater than $\xi$) and $\bar v$ is the r.m.s string 
velocity.  The coherence time scale in a Fourier mode of wavenumber $k$ is
determined  by the time segments take to travel a distance $k^{-1}$ [8].
We find that the characteristic coherence time scale for a mode of spatial
frequency $k$ is $\eta_c \simeq 3/k$ [8].  Our numerical results indicate that
at high $k$, $\eta_c$ decreases faster than $k^{-1}$, but we believe that this
behaviour is a lattice artifact.  The implications of our 
simulations for the appearance of the Doppler peaks are not entirely clear 
cut.  The coherence time is smaller than, but of the same order of magnitude 
as, the period of acoustic oscillations in the photon baryon fluid at 
decoupling, which is roughly $11/k$ [9].  This is in turn smaller than the 
time at which the power in the energy and velocity sources peak, 
approximately $20/k$ [8].  We believe that our string correlation functions
can serve as realistic sources to answer the question of existence or
absence of secondary peaks in the CMB angular power spectrum.

\section {References}

\noindent
[1] P. J. Steinhard, {\sl Class. \&  Quant. Grav.} {\bf10} (1993) S33.

\noindent
[2] R. Durrer, A. Gangui, and M. Sakellariadou,
{\sl Phys. Rev. Lett.} {\bf76} (1996) 579.

\noindent
[3] R. Durrer and Z.H. Zhou, {\sl Phys. Rev.} D{\bf 53} (1996) 5394.

\noindent
[4] R.G. Crittenden and N. Turok, {\sl Phys. Rev. Lett.} {\bf75} (1995) 2642.

\noindent
[5] A. ALbrecht, D. Coulson, P. Ferreira and J. Magueijo, 
{\sl Phys. Rev. Lett.} {\bf76} (1996) 1413.

\noindent
[6] J. Magueijo, A. Albrecht, P. Ferreira and D. Coulson,
``The structure of Dopple rpeaks induced by active perturbations'',
astro-ph/9605047 (1996).

\noindent
[7] R. Durrer and M. Sakellariadou (in preparation).

\noindent
[8] G. Vincent, M. Hindmarsh and M. Sakellariadou,
``Correlations in cosmic string networks'',
astro-ph/9606137 (1996).

\noindent
[9] P.J.E. Peebles, ``{\sl The Large Scale Structure of the Universe}''
(Princeton University Press, Princeton, 1980).

\end